\documentclass{pasj00}
\begin{document}
\SetRunningHead{K. Sadakane et al.}{Na~{\sc i} Lines in V1280 Sco}
\Received{}
\Accepted{}

\title{Discovery of Multiple High-Velocity  Narrow Circumstellar Na~{\sc i} D Lines in Nova V1280 Sco
\thanks{Based on data collected at the Subaru Telescope,
                  which is operated by the National Astronomical Observatory
of Japan.}} 

\author{Kozo \textsc{Sadakane},\altaffilmark{1}
        Akito \textsc{Tajitsu},\altaffilmark{2} 
        Sahori \textsc{Mizoguchi},\altaffilmark{3} 
        Akira \textsc{Arai},\altaffilmark{4} 
        and        
        Hiroyuki \textsc{Naito},\altaffilmark{5}
       }

\altaffiltext{1}{Astronomical Institute, Osaka Kyoiku University, Asahigaoka,  Kashiwara-shi, Osaka
         582-8582}
\altaffiltext{2}{Subaru Telescope, National Astronomical Observatory of Japan, 650 North A'ohoku Place, 
Hilo, HI 96720, USA}
\altaffiltext{3}{Sendai Astronomical Observatory, Nishikigaoka, Aoba-ku, Sendai-shi, Miyagi, 989-3123}
\altaffiltext{4}{Department of Physical Science, Hiroshima University, 1-3-1 Kagamiyama, Higashi-Hiroshima-shi,
Hiroshima 739-8526}
\altaffiltext{5}{Department of Physics, Nagoya University, Furo-cho, Nagoya-shi, Aichi  
464-8602}

\email{sadakane@cc.osaka-kyoiku.ac.jp}


\maketitle

\begin{abstract}

We discovered multiple high-velocity (ranging from -900 to -650 km 
s${}^{-1}$) and  narrow (FWHM $\sim$ 15 km s${}^{-1}$) absorption 
components corresponding to both the D2 and the D1 lines of Na~{\sc i} on 
a high dispersion spectrum of V1280 Sco observed on 2009 May 9 (UT), 
814 d after the $\it V$ band maximum.  Subsequent observations 
carried out on 2009 June  and July  confirmed at least 11 distinct  
absorption components in both systems. Some components had deepened 
during the two months period while their HWHMs and wavelengths 
remained nearly constant. 
We suggest these high velocity components originate in cool clumpy gas 
clouds moving on the line of sight, produced in interactions between 
pre-existing cool circumstellar gas and high velocity gas ejected in 
the nova explosion. 
The optical region spectrum of V1280 Sco in 2009 is dominated by the
continuum radiation and exhibits no forbidden line characterizing the
nebular phase of typical novae. Permitted Fe~{\sc ii} lines show doubly
peaked emission profiles and some  strong  Fe~{\sc ii} lines are 
accompanied by a  blue shifted ($\sim$ -255 km s${}^{-1}$) absorption 
component. However, no high-velocity and narrow components corresponding
 to those of Na~{\sc i} could be detected in Fe~{\sc ii} lines nor in 
the Balmer lines. The 255 km s${}^{-1}$ low velocity absorption component
 is most probably originating in the wind from the nova. 

\end{abstract}

\section{Introduction}

The bright nova V1280 Sco (Nova Sco 2007 $\sharp$1, \timeform{16h57m40s.91}, 
\timeform{-32D20'36''.4}, equinox 2000.0)  was discovered on 2007 February 4 
by two Japanese observers (Y. Sakurai and Y. Nakamura) independently,
as  communicated by \citet{yamaoka2007}. 
\citet{naito2007} obtained a low dispersion spectrum of the object on Feb. 
5.87 and found that it had a smooth continuum together with the Balmer and 
Fe~{\sc ii}  lines showing P Cygni profiles. 
The nova had brightened greatly during the first 10 days and reached the 
maximum brightness $\it V$ = 3.8 on Feb. 16 \citep{munari2007a}. 
The outburst amplitude ($\it A$) is 15 mag or larger because \citet{das2008}
noted that no star was visible down to $\it B$ and $\it R$ magnitudes of 20.3
and 19.3, respectively, on pre-discovery plates. 
\citet{munari2007a} reported their 
photometric and spectroscopic monitoring of the nova and noted that on Feb. 20.24 
the nova showed a typical classical nova spectrum of the Fe~{\sc ii} type 
characterized by a rich forest of strong permitted emission lines of Fe~{\sc ii} 
displaying deep P Cyg profiles.

\citet{das2007} observed the nova in the near IR region on 2007 March 4.95 and 
found that the continuum in the 1.08 - 2.35 $\mu$ region had risen sharply 
indicating the  dust formation in the nova ejecta. \citet{puetter2007} reported 
spectroscopic observations in the visual - near IR regions carried out in 2007 
May and found that the nova was in a very low excitation state showing strong 
C~{\sc i} lines and no discernible He~{\sc i} emission.
They estimated the reddening, which is in part due to the dust shell, from the 
O~{\sc i} lines and derived E($\it B - V$) = 1.7 mag. Photometric behaviors of the 
nova in the optical  region  until 2007 October 9 are summarized by 
\citet{munari2007b}. They pointed out a re-brightening in 2007 May
and noticed an unexpected large re-brightening in 2007 September.

High spatial resolution interferometric observations in the near and mid IR 
regions were carried out using the Very Large Telescope Interferometer (VLTI). 
\citet{chesneau2008} observed spectra and visibilities of the nova from 2007 
February 28 to June 30 and determined an apparent linear expansion rate for the dust 
shell. They  pointed out that the approximate time of the mass ejection during which the dust 
shell had formed was close to the date of the maximum brightness.  They used the observed 
expansion velocity and the linear expansion rate to derive an upper limit of the 
distance to be  $\sim$ 1.9 kpc.

We have continued photometric observations in $\it B, V, Rc, Ic$ and $\it y$ bands at 
Osaka Kyoiku University and in $\it J, H$ and $\it Ks$ bands at Higashi Hiroshima
Observatoty until 2009 September. Low resolution spectroscopic observations were 
conducted  at Nish-Harima Astronomical Observatory. Part of our photometric and
spectroscopic observations has been reported in \citet{naito2009}. A long-term
(spanning three years) light curve of V1280 Sco reveals unique features among 
classical novae.
Figure 1 shows long-term light curves in the $\it V$ and the $\it y$ bands including 
the latest observations obtained in the summer of 2009.
After the initial rapid decline in the $\it V$ band, the nova showed a 
temporary re-brightening in 2007 May. Then it declined and reached a minimum
($\sim$ 16 mag in the $\it V$ band) in 2007 August. After 2007 September, the nova
recovered its brightness in all bands and a bright plateau ($\sim$ 10 mag in the 
$\it V$ band)  had continued throughout 2008. Although the nova showed a
slight decline starting from 2009 May, it was still brighter than 11 mag in the
 $\it V$ band in 2009 August, more than 900 d after the  $\it V$ band maximum.
One of the notable feature of the optical light curve is a close resemblance 
between the $\it y$, which is free from emission lines,
 and the $\it V$ magnitudes. Even in the summer of 2009,
its $\it y$ magnitude is only slightly fainter than the $\it V$ magnitude.
This implies that the optical region spectrum is dominated by the continuum
 radiation  even 900 d after the $\it V$ band maximum. 
A detailed discussion of the light and color curves will be presented in a
separate paper.
A low resolution 
optical spectrum of V1280 Sco obtained at Higashi Hiroshima Observatoty 
on 2009 February shows no hint of the [O~{\sc iii}] forbidden lines. Thus, 
the nova has not entered the Nebular phase yet. 
\begin{figure}[t]
     \begin{center}
       \FigureFile(75mm,70mm){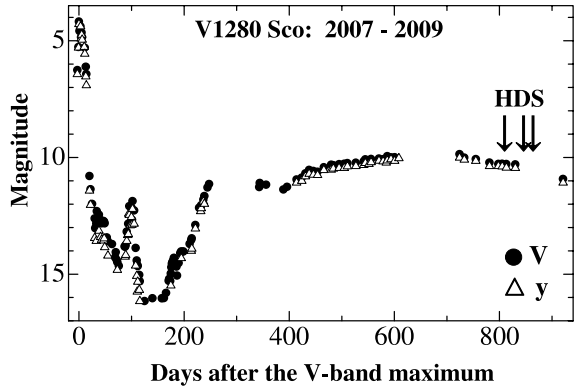}
     \end{center}
     \caption{Light curves in the $\it V$ (dots) and the $\it y$ (open
              triangles) bands obtained at Osaka Kyoiku University.
              Data of photometric starndard stars given by \citet{henden2007}
              are used. Epochs of Subaru observations in 2009 are indicated by arrows.
               }
    \end{figure}
We have carried out high resolution optical region spectroscopic observations
of the nova in 2009 in order to clarify the physical conditions in the
continuum radiating source. Here we focus on  a remarkable finding concerning 
the highly blue-shifted narrow absorption components associated with the
Na~{\sc i} yellow doublet lines. 

\section{Observations}

Spectroscopic observations of V1280 Sco were carried out at three 
epochs with the Subaru Telescope using the High Dispersion Spectrograph (HDS) 
on 2009 May 9, 
 June 15 and 16 and July 4 and 6 (UT), as summarized in table 1 . 
Our first observation was made 
814 d after the optical maximum. On May 9, we obtained a sky
observation using the same instrumental setup because the sky was hazy
 and the object was located close to the bright Moon. 
The sky condition on the subsequent nights in June and July was fairly clear
and  no sky subtraction had been applied for data obtained in June and July. 
Technical details
and the performance of the spectrograph are described in \citet{nogu02}. We used
a slit width of \timeform{0".6} (0.3 mm) and a 2x1 binning mode, which enabled
us to achieve a nominal spectral resolving power of about $R = 60000$ with
a 3.5 pixel sampling. Our observations covered the wavelength region from 
4050 \AA~ to 6760 \AA. For flat-fielding of the
CCD data, we obtained Halogen lamp exposures (flat images) with the same setup
as that for the object frames.
\setcounter {table} {0}
\begin{table}
      \caption{Log of spectroscopic observations .}\label{first}
\footnotesize
      \begin{center}
      \begin{tabular}{cccc}
\hline\hline
 Date   & UT  & Day  & Exposure  \\
 2009   &     & after the $\it V$max & sec \\
\hline
May 9   & 12:30	& 814	& 900    \\
June 15 & 12:30	& 851	& 600    \\
June 16 & 11:55	& 852	& 900    \\
July 4  & 11:00	& 870	& 900    \\
July 6  & 9:30	& 872	& 1200    \\

\hline
        \end{tabular}
      \end{center}
\end{table}

The reduction of two-dimensional echelle spectra was performed using the
IRAF software package in a standard manner. The wavelength calibration was performed using the
Th-Ar comparison spectra obtained during the observations. Extracted 
one dimensional spectral data have been converted to the helio-centric scale 
and then continuum fitting was done using high order polinomials. 
The spectral data obtained on June 15 and 16
and July 4 and 6 were averaged to create data of June and July, respectively, 
in order to improve the signal-to-noise (SN)  ratio. The measured FWHM of
the weak Th lines was 0.11 \AA~ at 6000 \AA, and the resulting resolving power was
around $R = 55000$. Measured  SN ratios at 5000 \AA~ were $\sim$ 55, 140, and
130 for data obtained on May, June, and July, respectively.
The averaged FWHM of the atmospheric water vapor absorption lines near the Na~{\sc i} 
D lines is 5.6 $\pm$ 0.8 km s${}^{-1}$, confirming the above resolution. 
Figure 2 displays a flux calibrated  HDS spectrum of V1280 Sco obtained 
in June. We use the spectrum of an A0 III star HIP 83740 observed on June 15 and 16
with the same instrumental setup in the process of flux calibration.
\begin{figure}
     \begin{center}
       \FigureFile(75mm,70mm){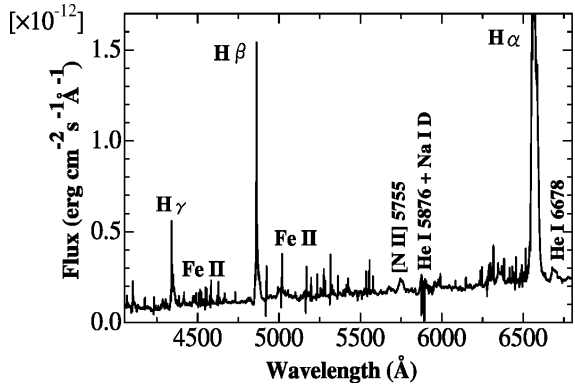}
     \end{center}
     \caption{Flux calibrated spectrum of V1280 Sco observed in 2009 June.
              The flux calibration might be somewhat uncertain because of the 
              variable seeing and the atmospheric transparency. 
              Major spectral features are indicated.  Sharp emission
              lines below 5600 \AA~ are identified as permitted lines of Fe~{\sc ii}.   }
    \end{figure}

\section{Results}

As a first step, we searched for signature of forbidden emission lines which are often
observed in  spectra of late-time novae. The most prominent features observed 
during classical novae nebular phase are the forbidden lines of doubly ionized oxygen, [O~{\sc iii}], 
at 4959.91 and 5006.84 \AA.  Figure 3 shows the region between 4955 \AA~ and 5025 \AA~
 observed on June, where the [O~{\sc iii}] lines are expected to be seen at positions
indicated by upward arrows. The figure shows that the spectrum is dominated by the continuous
radiation. We can find no trace of the emission lines on the high resolution
and high SN data. Instead, we find emission lines of Fe~{\sc ii}. The weak Fe~{\sc ii} line at
4993.358 \AA~ shows a double-peaked profile,  while the stronger line at 5018.440 \AA~ is
single-peaked. Interestingly, the latter line is accompanied by a blue-shifted ($\sim$
- 255 km s${}^{-1}$) absorption component. The same is true for another strong 
Fe~{\sc ii} line at 4923.927 \AA.  
\begin{figure}
     \begin{center}
       \FigureFile(80mm,100mm){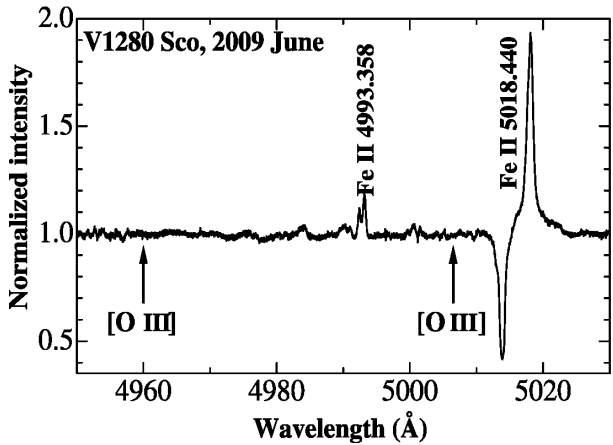}
     \end{center}
     \caption{Absence of the [O~{\sc iii}] forbidden lines. Expected positions
              are indicated by upward arrows. Two prominent emission lines 
              of Fe~{\sc ii} are labeled.}
    \end{figure}

Inspecting the spectral region near the Na~{\sc i}  doublet lines, we noticed unfamiliar
sharp absorption lines between 5870 and 5885 \AA. These sharp features are superposed on
a wide (FWHM $\sim$ 1300 km s${}^{-1}$) emission line of He~{\sc i} at 5875.62 \AA.
We trace the profile of the emission line and use the result as a $\it pseudo$ continuum
in the present study. Resulting normalized spectrum is shown in figure 4. The two strong
lines labeled as D2 and D1 are most probably interstellar (IS) absorptions. 
At least six IS components can be recognized on the data. 
The averaged FWHM of weak interstellar lines is 9.0 $\pm$ 1.0  km s${}^{-1}$. 
We notice that the pattern of
absorption features between 5872 and 5878 \AA~ is just the same with the pattern
observed between 5878 and 5884 \AA. We measured the differences in wavelength between the
former pattern and the strongest IS D2 line and those between the latter pattern and 
 the strongest IS D1 line to find that the differences are exactly the same for
all corresponding features.
Thus, we conclude that the unfamiliar sharp  lines are blue-shifted absorptions of 
both D2 and D1 lines.
\begin{figure}
     \begin{center}
       \FigureFile(75mm,130mm){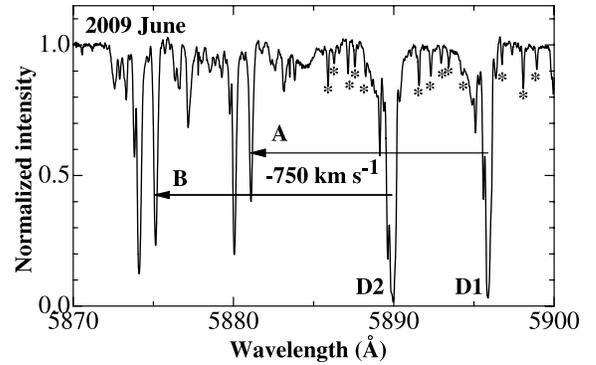}
     \end{center}
     \caption{Normalized spectral data around the D2 and the D1 lines of Na~{\sc i}.
              Interstellar absorptions are labeled as D2 and D1. Arrows A and B indicate
              the blueward shifts of both the D2 and the D1 systems, respectively. Weak absorption
              lines labeled with asterisks are due to atmospheric water vapor.}
    \end{figure}
\begin{figure}
     \begin{center}
       \FigureFile(75mm,70mm){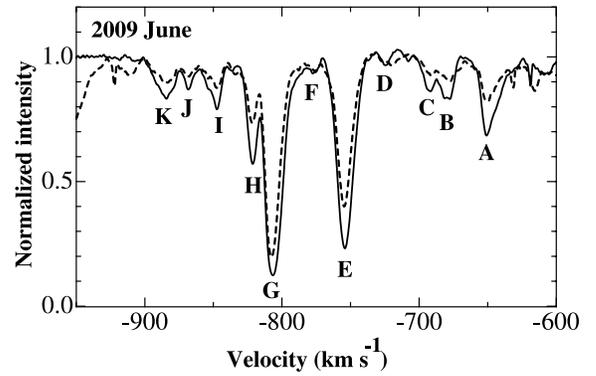}
     \end{center}
     \caption{Blue shifted D2 (thick) and D1 (dashed) absorption lines observed on
              2009 June plotted on the velocity scale.
              Eleven individual components are labeled from A to K.}
    \end{figure}
\begin{figure}
     \begin{center}
       \FigureFile(75mm,70mm){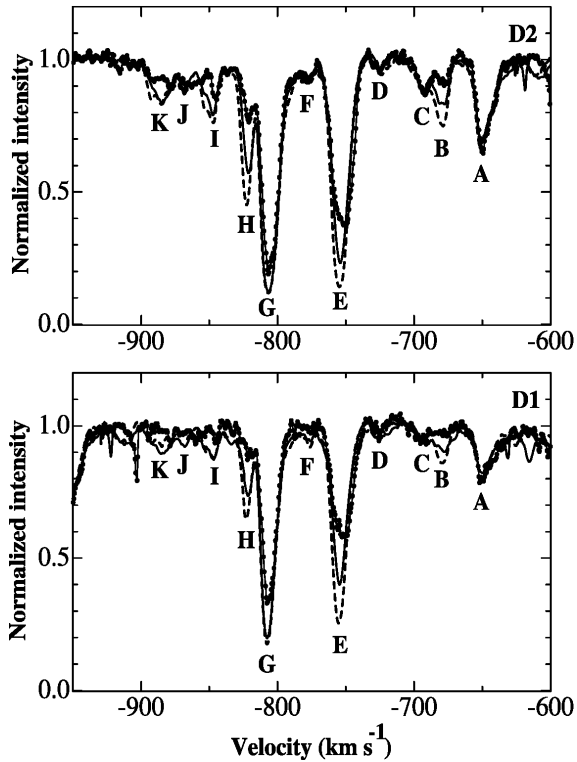}
     \end{center}
     \caption{Changes in the blue shifted absorptions in two months.
              Data observed on May, June, and July are displayed by dotted,
              thick and dashed lines, respectively. The D2 and the D1 systems 
              are displayed in the upper and the lower panels, respectively.  }
    \end{figure}


 We converted the wavelength scale to the velocity scale 
originating from the laboratory wavelengths of both D2 and D1 lines. 
Figure 5 displays blue-shifted absorption features observed on 2009 June plotted
on the velocity scale. 
The two patterns which originate from the D2 and the D1 lines coincide exactly. 
We can recognize at least 11 pairs of absorption features and they are labeled as 
components A to K on the figure. The features in the D1 
absorption system are systematically weaker than those in the D2 system. 
The difference reflects the values of the transition probabilities of the D2 and 
the D1 lines.

We examine variations in strengths of the above 11 absorption features during the
2 months period (from May to July) in figure 6. Lines belonging to the D2 and the
D1 systems are displayed separately on the figure. 
Measurements of the velocities and the depths for the 11 components belonging to
the D2 system are given in table 2. Errors in measurements are estimated to be
$\pm$ 2.5 km s${}^{-1}$ in the velocity and $\pm 0.03$ in the depth.
We find definite increase 
in line depth for six components (B, E, G, H, I, and K) in both the D2 and the D1
systems. Depths of the remaining five components stay nearly the same during the
period. It is interesting to find that no component has weakened during the 2 
months period. 
The strongest feature G had deepened between May and June, while
its depth remained constant between June and July. The neighboring component H
had significantly deepened from $\it r$ = 0.24 in May to 0.55 in July. 
The two neighboring components B and C clearly show different behaviors during  
the two months period. The former had deepened significantly (from $\it r$ = 0.10 to 0.24),  
while the depth of the latter
remained nearly constant. The averaged FWHM of the three apparently unblended absorptions
(A, E, and G) is 16.0 $\pm$ 2 km s${}^{-1}$, while that  of the narrowest feature 
(component H) is 11.2 km s${}^{-1}$, nearly the same as that of the weak interstellar
absorptions. 
\setcounter {table} {1}
\begin{table}
      \caption{Measurements of the D2 system.}\label{second}
\footnotesize
      \begin{center}
      \begin{tabular}{ccccccc}
\hline\hline
 Component   & & Velocity & & & Depth &   \\
             & & km s${}^{-1}$ & & & &    \\
             & May & June & July & May & June & July \\
\hline
A & -649.4 & -650.4 & -650.4 & 0.35 & 0.32 & 0.31 \\
B & -677.9 & -679.0 & -680.0 & 0.10 & 0.16 & 0.24 \\
C & -692.2 & -692.2 & -692.0 & 0.13 & 0.14 & 0.12 \\
D & -724.3 & -724.5 & -724.4 & 0.04 & 0.04 & 0.06 \\
E & -751.8 & -753.8 & -754.8 & 0.62 & 0.76 & 0.86 \\
F & -776.8 & -776.2 & -776.8 & 0.08 & 0.06 & 0.08 \\
G & -806.3 & -806.3 & -806.8 & 0.80 & 0.87 & 0.88 \\
H & -821.1 & -821.6 & -822.6 & 0.24 & 0.43 & 0.55 \\
I & -845.5 & -847.5 & -848.1 & 0.14 & 0.21 & 0.23 \\
J & -867.0 & -867.9 & -867.9 & 0.11 & 0.13 & 0.12 \\
K & -885.8 & -884.7 & -885.8 & 0.08 & 0.17 & 0.17 \\
\hline
        \end{tabular}
      \end{center}
\end{table}
It is interesting to examine whether the highly blue-shifted absorptions can be
seen associated with other metal lines or the Balmer emission lines. 
A region near the strong Fe~{\sc ii} line 5018.440 \AA~ is displayed in figure 7
and the profile of H$\beta$ is shown in figure 8.
Expected positions for the two strong absorption features (E and G) are indicated
by arrows. We can confirm no  absorption feature in these lines and thus 
conclude that the highly blue-shifted absorptions are only associated with the resonant 
transitions of Na~{\sc i}.   A relatively broad (HWHM $\sim$ 60 km s${}^{-1}$) absorption
feature is seen in figures 7 and 8, which is blue-shifted by $\sim$ -255 km s${}^{-1}$
relative to the emission peak. We interpret this absorption originates in the expanding 
wind of the photosphere. 
\begin{figure}
     \begin{center}
       \FigureFile(70mm,70mm){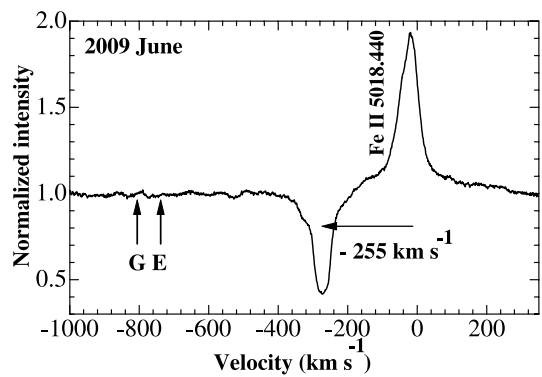}
     \end{center}
     \caption{Absence of the highly blue shifted absorptions in the strong Fe~{\sc ii}
              line. Expected positions of the two strong components E and G  of Na~{\sc i}
              are indicated by upward arrows.}
    \end{figure}
\begin{figure}
     \begin{center}
       \FigureFile(70mm,70mm){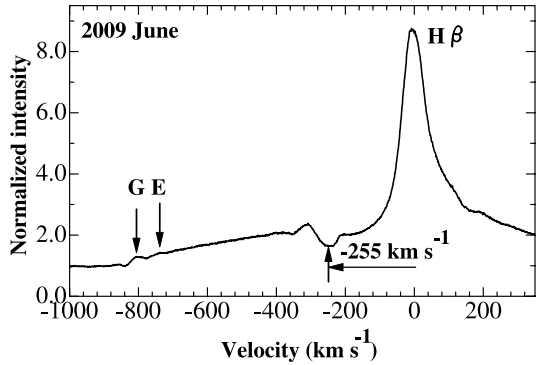}
     \end{center}
     \caption{Absence of the highly blue shifted absorptions in H$\beta$. 
     Expected positions of the two strong components E and G of Na~{\sc i} are
              indicated by downward arrows.}
    \end{figure}

\section{Discussion}

We have discovered multiple high-velocity absorption components corresponding to 
both the D2 and the D1 lines of Na~{\sc i} on  high resolution spectra of 
V1280 Sco observed three years after the explosion. There are
at least 11 sharp (FWHM $\sim$ 15 km${}^{-1}$) components and they are highly blue-shifted, 
ranging from -900 to -650 km s${}^{-1}$, on the helio-centric velocity scale. 
These absorptions are not associated with Fe~{\sc ii} or the Balmer emission lines.  
This implies that the   high-velocity absorptions  are produced in a cool environment
where iron atoms are in neutral state and most hydrogen  atoms stay in the ground state.  
This must be a very rare finding because we know no previous similar observation in the
literature. 

\citet{williams2008} reported observations of short-lived blue-shifted metallic 
absorption systems, including the Na~{\sc i} D lines, near the maximum light of various novae.
These absorption lines have expansion velocities from 400 to 1000 km s${}^{-1}$ and velocity
dispersions between 35 and 350 km s${}^{-1}$. They are usually accelerated outward  
and progressively weaken and disappear over timescales of weeks (within 100 d). 
They proposed a spiral ring model to interpret their observations.    They suggest that
some material ejected by the secondary star before the nova outburst is spiraling around the
binary system. After the nova outburst, a rapidly expanding luminous photosphere is 
produced and it collides with the pre-existing gas stream. 

Our observations of narrow and high-velocity absorption components associated with the 
Na~{\sc i} D lines appear hard to be interpreted in this scheme. Our data were obtained
more than 800 d after the explosion, while the transient absorptions reported in \citet{williams2008} 
usually disappear over timescales of weeks. Their analysis of absorption line systems observed
in LMC 2005 shows the excitation temperature to be around 10${}^{4}$ K, which is too 
high to explain the absence of high-velocity absorption components associated with Fe~{\sc ii} 
lines. Furthermore, the observed line widths of the high-velocity absorption components in V1280 Sco
are narrower than the widths noted in  \citet{williams2008}.  Finally, 
the large number (at least 11) of the absorption components looks difficult to 
be interpreted in this scenario.

We propose that the observed high velocity absorption components originate in many cool
 clumpy gas clouds moving toward the observer on the line of sight.
These gas clouds are produced after the nova explosion in interactions between the
pre-existing cool circumstellar gas and the high velocity gas ($\sim$ 2000 km s${}^{-1}$,
Naito et al. 2009)  ejected in the nova explosion. 
\citet{naito2009} suggested that the
2000 km s${}^{-1}$ gas might be associated with the second mass ejection episode
corresponding to the temporal re-brightening observed in 2009 May.
In the case of V1280 Sco, the binary system had been 
embedded in a relatively dense circumstellar cloud, and the pre-existing circumstellar gas 
had not necessarily been supplied from the secondary star as suggested in \citet{williams2008}. 
 The prompt formation of dust in this nova \citep{chesneau2008} is likely 
to be triggered by the interaction between the surrounding clouds and the expanding ejecta. 
Many small gas clouds might have been produced during the turbulent interaction in the shock 
and expelled outward with velocities ranging from -1000 to -500 km s${}^{-1}$.  
We suppose that some 
of the clouds are moving toward us on the line-of-sight and produce multiple absorption 
lines. The observed increase in line depths for several components may be the result of
rapid cooling of these clouds, increasing the number density of neutral Na capable of 
absorbing the D line photons. This picture is hinted by a direct image of the old nova 
GK Per observed with the WIYN 3.5 m telescope \citep{slavin1995}. 
Confirmation of this scenario could come from future high spatial resolution imaging observations.

\vskip 3mm
We thank Dr. M. Kato for providing a $\it y$ band filter and Dr. 
I. Hachisu for helpful comments and suggestions. 
Co-operations of students at Osaka Kyoiku University are gratefully acknowledged.
This research was partly supported by Grants-in-Aid from the Ministry of 
Education, Culture, Sports, Science and Technology (No. 19540240 KS).



\begin{thebibliography}{}

\bibitem[Chesneau et al. (2008)] {chesneau2008}
        Chesneau, O., et al. \ 2008, \aap, 487, 223


\bibitem[Das et al. (2007)] {das2007}
        Das, R. K., Ashok, N. M.,  \& Banerjee, D. P. K. 
         \ 2007, Central Bureau Electronic Telegrams, 866 

\bibitem[Das et al. (2008)] {das2008}
        Das, R. K., Banerjee, D. P. K., Ashok, N. M., \& 
        Chesneau, O. \ 2008, \mnras, 391, 1874




\bibitem[Henden and Munari (2007)]{henden2007}
       Henden, A., \& Munari, U. \ 2007, IBVS, 5771



\bibitem[Munari et al. (2007b)] {munari2007b}
        Munari, U., Siviero, A., Henden, A., Ochner, P., Simoncelli, C.,
        Tomasoni, S., Moschini, F., \& Dallaporta, S.
        \ 2007b, Central Bureau Electronic Telegrams, 1099 

\bibitem[Munari et al. (2007a)] {munari2007a}
        Munari, U., Valisa, P, Dalla Via G., \& Dallaporta, S.
        \ 2007a, Central Bureau Electronic Telegrams, 852 


\bibitem[Naito and Narusawa (2007)]{naito2007}
         Naito, H., \& Narusawa, S. \ 2007, IAU Circ., 8803-2

\bibitem[Naito et al. (2009)]{naito2009}
         Naito, H., Mizoguchi, S., Arai, A., Yamanaka, M., Narusawa, S., Sadakane, K., and
         Iijima, T. \ 2009, in Proc. 10th Asian-Pacific Regional IAU Meeting, ed. S. N. Zhang, 
         Y. Li, \& Q. Yu (Beijing, China Science \& Technology Press), 246

\bibitem[Noguchi et al.(2002)]{nogu02}
         Noguchi, K., et al.  \ 2002,   \pasj, 54, 855


\bibitem[Puetter et al. (2007)]{puetter2007}
         Puetter, R. C., Rudy, R. J., Lynch, D. K.,  Russell, R. W., 
         Mazuk, S., Pearson, R. L., Woodward C. E., \& Perry, R. B. \ 2007, 
         IAU Circ., 8845-1


\bibitem[Slavin et al. (1995)]{slavin1995}
         Slavin, A. J., O'Brien, T. J., \& Dunlop, J. S. \ 1995, 
        \mnras, 276, 353


%

\bibitem[Williams et al. (2008)]{williams2008}
       Williams, R., Mason, E., Della Valle, M., \& Ederoclite, A. \ 2008, 
       \apj, 685, 451


\bibitem[Yamaoka et al. (2007)]{yamaoka2007}
         Yamaoka, H., Nakamura, Y., Nakano, S., Sakurai, Y., \& Kadota, K.
         \ 2007, IAU Circ., 8803-1





\end{thebibliography}
\end {document}